# ARTIFICIAL NEURAL NETWORKS AS NON-LINEAR EXTENSIONS OF STATISTICAL METHODS IN ASTRONOMY

Ofer Lahav

Institute of Astronomy, Madingley Road

Cambridge CB3 0HA, UK

e-mail: lahav@mail.ast.cam.ac.uk

**Abstract** — We attempt to de-mistify Artificial Neural Networks (ANNs) by considering special cases which are related to other statistical methods common in Astronomy and other fields. In particular we show how ANNs generalise Bayesian methods, multi-parameter fitting, Principal Component Analysis (PCA), Wiener filtering and regularisation methods. Examples of morphological classification of galaxies illustrate how non-linear ANNs improve on linear techniques.

## 1. Introduction

Artificial Neural Networks (ANNs) have recently been utilised in Astronomy for a wide range of problems, e.g. from adaptive optics to galaxy classification (for review see Miller 1993 and Storrie-Lombardi & Lahav 1994). While ANNs seem to be practically useful, it has little been discussed in the Astronomical literature how they are related to other statistical methods. Questions commonly asked by 'Neuro-sceptics' are:

- Could we understand what the ANNs are doing, or are they just 'black boxes'?

- If one has already selected 'good parameters', does it matter what classifier is to be used ?

Here we attempt to show that the ANNs approach should be viewed as a general statistical framework, rather than as an esoteric approach. It is shown that some special cases of ANNs are statistics we are all familiar with. However, the ANNs can do better, by allowing non-linearity. There is of course freedom in choosing what kind of 'non-linearity' to apply, but sensible choices show that significant improvement can be achieved over the linear approaches. Here we illustrate these points by some examples from the problem of morphological classification of galaxies, using the ESO-LV (Lauberts & Valentijn 1989) sample with 13 parameters and 5217 galaxies, as analysed by ANNs in Storrie-Lombardi et al. (1992) and Lahav et al. (1995). The latter paper also gives more mathematical details on the issues discussed below.

For cosmologists, there is an analogy here with $N$-body simulations of gravitational systems. Linear theory is reasonably well understood, but is not sufficient to describe complicated dynamics. One needs to use then numerical simulations, producing results which are not always understood by intuition or by analytic methods. However, one can verify what is happening by considering simple cases (e.g. the spherical infall model) to gain confidence in what the simulations give. Our approach to the ANNs is similar.

## 2. ANNs as Minimization Algorithms

It is very common in Astronomy to fit a model with several (or many) free parameters to the observations. This regression is usually done by means of $\chi^2$ minimization. A simple example of a 'model' is a polynomial with the coefficients as the free parameters. Consider now the specific problem of morphological classification of galaxies. If the type is $T$ (e.g. on de Vaucouleurs' numerical system [-6,11]) and we have a set of parameters $\mathbf{x}$ (e.g. isophotal diameters and colours) then we would like to find free parameters $\mathbf{w}$ ('weights') such that the 'cost function'

$$E = \frac{1}{2}\sum_i [T_i - f(\mathbf{w}, \mathbf{x_i})]^2, \qquad (1)$$

where the sum is over the galaxies, is minimized. The function $f(\mathbf{w}, \mathbf{x})$ is the 'network'. Commonly $f$ is written in terms of

$$z = \sum_k w_k x_k, \qquad (2)$$

where the sum here is over the input parameters to each node. A 'linear network' has $f(z) = z$, while a non-linear transfer function could be a sigmoid $f(z) = 1/[1+\exp(-z)]$ or $f(z) = \tanh(z)$. Another element of non-linearity is provided by the 'hidden-layers'. The 'hidden layers' allow curved boundaries around clouds of data points in the parameter space. A typical configuration with one 'hidden-layer' and a single output for the galaxy type $T$ is shown in Figure 1.

While in most computational problems we only have 10-1000 nodes, in the brain there are $\sim 10^{10}$ neurons, each with $\sim 10^4$ connections. For a given Network architecture the first step is the 'training' of the ANN. In this step the weights are determined by minimizing 'least-squares'. The Backpropagation algorithm (Rumelhart, Hinton & Williams 1986; Hertz, Krogh & Palmer 1991) is one of the most popular ANN minimization algorithms. However, there are other more efficient methods such as Quasi-Newton (e.g. Hertz et al. 1991).

The interpretation of the output depends on the network configuration. For example, a single output node provides an 'analog' output (e.g. for predicting the type or luminosity of a galaxy), while several output nodes can be used to assign probabilities to different classes (e.g. 5 morphological types of galaxies), as explained below.

## 3. The Perceptron as a Wiener Filter

The weights, the free parameters of the ANN, have a simple interpretation when the network is linear without hidden layers, commonly called the 'perceptron'. Let the input and output vectors be $\mathbf{x}$ and $\mathbf{s}$ respectively. The weights then form a matrix $W$ (not necessarily square), and the minimum variance $\langle (\mathbf{s} - W\mathbf{x})(\mathbf{s} - W\mathbf{x})^T \rangle$ with respect to the weights occurs for

$$W_{\text{opt}} = \langle \mathbf{s}\mathbf{x}^T \rangle \langle \mathbf{x}\mathbf{x}^T \rangle^{-1}. \tag{3}$$

This is in fact the standard Wiener (1949) filter known in digital filtering and image processing, commonly applied for signal+noise problems when $\mathbf{x} = \mathbf{s} + \mathbf{n}$ (e.g. Rybicki & Press 1992 for review, and Lahav et al. 1994a Fisher et al. 1994 and Zaroubi et al. 1994 for recent cosmological applications). We note that the same result can be derived by conditional probabilities with Gaussian probability distribution functions, as well as by regularisation with a quadratic prior.

One can go one step further, to generalize the above to non-linear *input*. This can be done e.g. by expanding the elements of the input vector as products of their powers. For example, if the input parameters are $x_1$ and $x_2$ the expanded input vector is

$$[1, x_1, x_2, x_1^2, x_1 x_2, x_2^2, ...].$$

This is sometime called the Volterra Connectionist Model. Other possible non-linear operations on the input are 'radial basis function', spherical harmonics and other non-linear functions. In fact, this can be viewed as an ad-hoc hidden layer which forces the input to a new non-linear form. The advantages are that the network is then easy to implement and fast to train. Moreover, the global minimum is unique.

4. Bayesian Classification

A classifier can be formulated from first principles according to Bayes theorem:

$$P(T_j|\mathbf{x}) = \frac{P(\mathbf{x}|T_j)\ P(T_j)}{\sum_k P(\mathbf{x}|T_k)\ P(T_k)} \qquad (4)$$

i.e. the *a posteriori* probability for a class $T_j$ given the parameters vector $\mathbf{x}$ is proportional to the probability for data given a class (as can be deduced from a training set) times the *prior* probability for a class (as can be evaluated from the frequency of classes in the training set). However, applying eq. (4) requires parameterization of the probabilities involved. It is common, although not always adequate, to use multivariate Gaussian:

$$P(\mathbf{x}|T_j) = (2\pi)^{-M/2}\ |C_j|^{-1/2}\ \exp[-\frac{1}{2}\mathbf{x}^T\ C_j^{-1}\ \mathbf{x}], \qquad (5)$$

$\mathbf{x}$ is of dimension $M$ and here has zero-mean, $\mathbf{x}^T$ is its transposed vector, and $C_j = \langle \mathbf{x}^T\ \mathbf{x} \rangle_j$ is the covariance matrix *per class j*.

It can be shown that the ANN behaves like a Bayesian classifier, i.e. the output nodes produce Bayesian *a posteriori* probabilities (e.g. Gish 1990), although it does not implement Bayes theorem directly. It is reassuring (and should be used as a diagnostic) that the sum of the probabilities in an 'ideal' network add up approximately to unity. Moreover, if both the training and testing sets are drawn from the same parent distribution, then the frequency distribution $P(T_j)$ for the objects as classified by the ANN is similar to that of the training set. In the case of a sigmoid output, it can be shown that the argument of the sigmoid is modelling the log-likelihood ratio of the two classes. The link between minimum variance and probability also illustrates why a classification scheme where one calculates the Euclidean distance of the ANN output from the vector representing each of the possible classes and then assigns the object to the class producing the minimum distance is equivalent to assigning a class according to the highest probability. For more rigorous and general Bayesian approaches for modelling ANNs see MacKay (1992).

Our experiments with the ESO-LV galaxy data indicate that ANNs can achieve a better success rate than the Bayesian classifier with Gaussian probability functions (eqs. 4 & 5). For 5 broad classes (E, S0, Sa+Sb, Sc+Sd and Irr) the success rate for perfect match is 64 % using the non-linear ANN (with one hidden-layer and sigmoid functions), compared with only 56 % using the linear Bayesian classifier.

## 5. Regularisation and Weight Decay

As in other inversion problems, the determination of many free parameters, the weights $w_i$'s in our case, might be unstable. It is therefore convenient to regularise the weights, e.g. by preventing them from growing too much. In the ANN literature this is called 'weight decay'. This approach is analogous to Maximum Entropy, and can be justified by Bayesian arguments, with the regularising function acting as the prior in the weight-space. Note that this is a different application of Bayes theorem from the one discussed in the previous section, applied in the class-space.

One possibility is to add a quadratic prior and to minimize

$$E_{tot} = \alpha E_w + \beta E_D, \tag{6}$$

where $E_D$ is our usual cost function, based on the data, and

$$E_w = \frac{1}{2} \sum_{i=1}^{Q} w_i^2 \tag{7}$$

is the chosen regularising function, where $Q$ is the total number of weights. The coefficients $\alpha$ and $\beta$ can be viewed as 'Lagrange multipliers'. While sometime they are specified ad-hoc, it is possible to evaluate them 'objectively' by Bayesian arguments in the weight-space. This has been done in the context of ANNs by MacKay (1992), following earlier analysis in relation with Maximum Entropy by Gull(1989; see also Lahav & Gull 1989). It turns out that the number of 'well determined' weights can be deduced from the eigen-values of the Hessian $\beta \nabla \nabla E_D$, evaluated with the weights at which $E_{tot}$ is minimum. When all $Q$ weights are well-determined and the number of objects $N$ is much larger than $Q$ one finds

$$\alpha^{-1} \approx \sigma_w^2 = 2\hat{E}_w/Q = \sum w_i^2/Q , \tag{8}$$

and

$$\beta^{-1} \approx \sigma_D^2 = 2\hat{E}_D/N , \tag{9}$$

as expected for Gaussian probability distribution functions.

Using a network configuration 13:3:1 (with 46 weights, including 'bias') for the ESO-LV galaxy data, with both the input data and the output $T$-type scaled to the range [0, 1] and with sigmoid transfer functions (so all the weights are treated in the regularisation process on 'equal footing') we find $\frac{\alpha}{\beta} \approx 0.001$.

In this particular problem we find that while the weight decay stabilizes the results, it makes little difference to the resulting rms dispersion between the ANN and the expert's classification. With or without weight decay we get $\Delta T_{\rm rms} \sim 2.1$ (over the $T$-scale [-5, 11]).

We note that the addition of the regularisation term $E_w$ changes the location of the minimum, now satisfying $\nabla E_D = -\frac{\alpha}{\beta} \nabla E_w = -\frac{\alpha}{\beta} \mathbf{w}$ (where the last equality holds for eq. 7, reminding the force of harmonic oscillator). Therefore with regularisation the probability interpretation for the network's output (described in §4) is altered, and the Wiener solution is modified.

## 6. Neural PCA

A pattern can be thought of as being characterized by a point in an $M$-dimensional parameter space. One may wish a more compact data description, where each pattern is described by $M'$ quantities, with $M' \ll M$. This can be accomplished by Principal Component Analysis (PCA), a well known statistical tool commonly used in Astronomy (e.g. Murtagh & Heck 1987 and references therein). The PCA method is also known in the literature as Karhunen-Loéve or Hotelling transform, and is closely related to the technique of Singular Value Decomposition. By identifying the *linear* combination of input parameters with maximum variance, PCA finds $M'$ variables (Principal Components) that can be most effectively used to characterize the inputs.

PCA is in fact an example of 'unsupervised learning', in which an algorithm or a linear 'network' discovers for itself features and patterns (see e.g. Hertz et al. 1991 for review). A simple net configuration $M : M' : M$ (called 'encoder') with linear transfer functions allows finding $M'$ linear combinations of the original $M$ parameters. The idea is to force the output layer to reproduce the input layer, by least-squares minimization. If the number of 'neck units' $M'$ equals $M$ then the output will exactly reproduce the input. However, if $M' < M$, the net will find, after minimization, the optimal linear combination. By changing the transfer function from linear to non-linear (e.g. a sigmoid) one can allow 'non-linear PCA'. Serra-Ricart et al. (1993) have used the ESO-LV galaxy data described above and have compared standard PCA to non-linear encoder, illustrating how the latter successfully identifies classes in the data.

It is possible to design other special ANNs to extract Principal Components (Oja 1989). While these learning rules give insight to the link between PCA and ANN, it is easier in practice to extract the Principal Components by the standard method or by an encoder.

## 7. Discussion

We have shown that ANNs can be viewed as non-linear extensions of other well-known statistical methods in Astronomy. As with all statistical methods 'the proof of the pudding is in the eating', and conclusions on success or failure of methods do depend on the specific problem and the quality of the data. It is encouraging that in the problem of morphological classification of galaxies, one of the last remaining subjective areas in Astronomy, ANNs can replicate the classification by a human expert almost to the same degree of agreement as that between two human experts, to within 2 $T$-units (Lahav et al. 1994b). Some of the techniques described here have recently been applied to a new sample of 830 APM galaxies, as described by A. Naim in this volume and in Naim et al. (1995). The challenge for the future is to develop efficient methods for feature extraction and 'unsupervised' algorithms, where the data speak for themselves, without using prior expert's classification.

**Acknowledgements** – I thank my collaborators to the ANN galaxy classification project, A. Naim, L. Sodré & M. Storrie-Lombardi for their contribution to the work presented here, and K. Fisher, Y. Hoffman, C. Scharf & S. Zaroubi, for stimulating discussions on statistical inference in large scale structure problems.